\documentclass[10pt]{article}

\usepackage{listings}
\usepackage{graphicx}
\usepackage{psfrag}
\usepackage{amssymb}

\usepackage{caption}
\usepackage{subcaption}
\usepackage{amsthm}

\usepackage{cite}
\usepackage[square]{natbib}
\usepackage{url}

\newtheorem{theorem}{Theorem}

\title{Cayley Analysis of Mechanism Configuration Spaces using CayMos: 
Software Functionalities and Architecture 
\footnote{This research was supported in part by the research grant NSF CCF-1117695
and a research gift from SolidWorks.}}

\author{Menghan Wang, Meera Sitharam \footnote{University of Florida, Gainesville, FL, USA}\\
        Department of Computer \& Information Sciences \& Engineering\\
        University of Florida
}

\begin{document}

\maketitle

\begin{abstract}
{\it 
For a common class of 2D mechanisms called \emph{1-dof tree decomposable linkages},
we present a software \texttt{CayMos} 
which uses new theoretical results from \cite{Sitharam2011a,Sitharam2011b,sitharam2014beast} to implement efficient algorithmic
solutions for:
(a) meaningfully representing and visualizing the connected components in the Euclidean realization space; 
(b) 
finding a path of continuous motion  between two realizations in the same connected component,
with or without restricting the \emph{realization type} (sometimes called orientation type); 
(c) finding two ``closest'' realizations in different connected components. }

\end{abstract}

\section{Introduction}


A key underlying barrier to understanding underconstrained geometric
constraint systems is the classical problem of representing and efficiently finding 
 the \emph{Euclidean realization spaces} of
\emph{1-degree-of-freedom linkages}, or \emph{mechanisms} in 2D.
A {\em linkage} $(G,\bar{l})$, is a graph $G=(V,E)$ with fixed
length bars as edges, i.e. $\bar{l}: E \rightarrow \mathbb{R}$. 
A 2D \emph{Euclidean realization} or \emph{configuration} $G(p)$ of $(G,\bar{l})$ is 
an assignment of points $p: V \rightarrow \mathbb{R}^2$ to the vertices of $G$ satisfying the bar lengths in $\bar{l}$, 
modulo Euclidean motions. 
If the linkage is \emph{flexible} (i.e., the space of realizations of a
linkage is infinite), but the addition of a
single bar causes the linkage to become \emph{minimally rigid} 
(i.e. only finitely many realizations), then the linkage is called a 
\emph{mechanism} with \emph{1-degree-of-freedom (1-dof)}. 


Describing and analyzing realization spaces of 1-dof linkages in 2D is 
a  difficult problem with a long history. 
In fact, even for rigid linkages, the number of realizations can be exponential in the number of vertices
and not easy to estimate \citep{Borcea2004}. 
For flexible linkages, 
a well-known early result \citep{Kempe1875} 
shows that an arbitrary algebraic curve can be traced by the motion of a linkage joint. 
One outstanding example is the Peaucellier-Lipkin linkage, which transforms planar rotary motion into straight-line motion \citep{Kempe1877}. 
Versions of the problem play an important role in Computer-Aided-Design (CAD), 
robotics and molecular geometry \citep{Sacks2010,Ying1995,Sitharam2005}, but few results are known beyond individual or specific families of linkages. 

\noindent{\bf Note:} In the remainder of the paper, ``linkage" refers to ``2D linkage".

In this paper, we restrict ourselves to \emph{1-dof tree-decomposable linkages}.
The underlying graphs $G$ of such linkages 
are obtained by dropping an edge from so-called \emph{tree-decomposable graphs}. 
Tree-decomposable linkages are 
minimally rigid and well-studied, for example, in geometric constraint solving and CAD.
For tree-decomposable linkages, if the bar lengths $\bar{l}$ are in $\mathbb{Q}$, 
the coordinate values of a realization are solutions to 
a triangularized quadratic system with coefficients in $\mathbb{Q}$
(i.e. the coordinates of the realization belong to an extension field over $\mathbb{Q}$ obtained by nested square-roots). 
Such values are called ruler-and-compass solvable or \emph{quadratically-radically solvable} \emph{(QRS)} values.

A graph $G$ is 1-dof tree-decomposable if there exists a non-edge $f$, 
called a \emph{base non-edge} (there could be more than one), 
such that $G$ has the following graph theoretical construction from $f$:
at {\emph{Construction Step $k$}}, 
the graph constructed so far, $G_{k-1}$, is extended by
adding two new maximal tree-decomposable subgraphs, or \emph{clusters} $C_{k1}$ and $C_{k2}$ sharing a vertex $v_{k}$. 
In addition, $C_{k1}$ and $C_{k2}$ each has exactly one shared vertex, $u_k$ and $w_k$ respectively, 
with $G_{k-1}$.
A \emph{realization type} for a 1-dof tree-decomposable linkage 
specifies a \emph{local orientation} for the triple $(v_k,u_k,w_k)$,
for each construction step  $v_k \triangleleft (u_k,w_k)$. 
For a given realization type, 
a 1-dof tree-decomposable linkage $(G,\bar{l})$ has a simple ruler and compass realization $G(p)$
which parallels the graph theoretic construction. 
On the other hand, when a realization type is not specified, 
determining the existence of a realization is however NP-hard \citep{Sacks2010}.

\noindent\textbf{Note}: in this paper, we assume that the clusters are globally rigid, and 
we reduce clusters sharing only two vertices with the rest of the graph into edges.

\subsection{Problems with the state of the art} \label{sec:soa}

There are numerous existing 
softwares dealing with 1-dof tree-decomposable linkages, 
such as Geometry Expressions, SAM,  Phun, Sketchpad, Geogebra, D-cubed, 
etc. 
They have the following major functionalities: 
(\romannumeral 1) designing 1-dof tree-decomposable linkages for tracing out specific curves, 
especially by building new mechanisms based on a library of existing ones;
(\romannumeral 2) accepting user-specified parameters, ranges and realization types to generate 
continuous motion of the linkages. 

\smallskip
However, while some theoretical answers have been provided recently for the following issues (which will be described in the next subsection), until now there has been no software implementation that addresses these issues.

\noindent\textbf{(a)} How to canonically represent and visualize the connected components? 
In the above softwares, 
the realization space is typically represented as separate curves in 2D that are 
traced by each vertex of the linkage. 
In fact, 
a realization actually corresponds to a tuple of points, one on each of these curves.
I.e., the realization space is bijectively represented by a curve in the full 
\emph{ambient dimension} of $2|V|-3$ after factoring out rigid transformations, 
where $|V|$ is the number of vertices in the linkage. \\
\textbf{(b)} 
How to efficiently generate all connected components, and find a path of continuous motion between two arbitrary input realizations
in the same connected component? 
Until now, in order to generate continuous motion, 
the user must 
specify a range of a parameter containing the parameter value at the given realization.
Then either a single connected component is generated
for a subset of the specified range, 
or multiple segments of the realization space, under only the given realization type,  
are generated within the specified range. \\
\textbf{(c)} How to find the ``distance'' between connected components? 
Until now, 
for two realizations in different connected components, 
there is no software that finds how ``close" they can get towards each other by continuous motion, 
using a meaningful definition of ``distance" between connected components. 


\subsection{Previous work on Cayley configuration spaces of 1-dof 2D linkages}
\label{sec:previous_cayley}

The recent papers \cite{Sitharam2011a,Sitharam2011b,sitharam2014beast}
introduced the use of \emph{Cayley configuration space} to describe the 
realization space of a 1-dof, 2D linkage $(G,\bar{l})$.  
A Cayley configuration space is obtained by taking 
an {\em independent non-edge} $f$ with $G\cup f$ being minimally rigid, 
and asking for all possible lengths that $f$ can attain 
(\romannumeral 1) over all the realizations of $(G, \bar{l})$; 
(\romannumeral 2) over all realizations of $(G, \bar{l})$ of a particular realization type. 
For (\romannumeral 1) (resp. (\romannumeral 2)), each realizable length of $f$ is called a (resp. \emph{oriented}) \emph{Cayley configuration}, 
and the set of all such configurations is called the (resp. \emph{oriented}) \emph{Cayley configuration space} of the linkage $(G,\bar{l})$ on $f$, parametrized by the length of $f$. 
The Cayley configuration space is a set of disjoint closed intervals on the real line.
For a 1-dof tree-decomposable linkage $(G, \bar{l})$, 
any base non-edge $f$ can be taken as the Cayley non-edge parameter to give a reasonable Cayley configuration space. 

An important complexity measure of the Cayley configuration space 
is its \emph{Cayley complexity}, i.e. the algebraic descriptive complexity of the interval endpoints 
in the Cayley configuration space \citep{Sitharam2011a}.  
Specifically, if the endpoints are QRS values, the corresponding linkage is said to have \emph{low Cayley complexity}.
We observe that  many commonly studied mechanisms, including the Strandbeest, have low Cayley complexity.

\begin{figure*}[hbtp]
\begin{center}
\includegraphics[width=.85\linewidth]{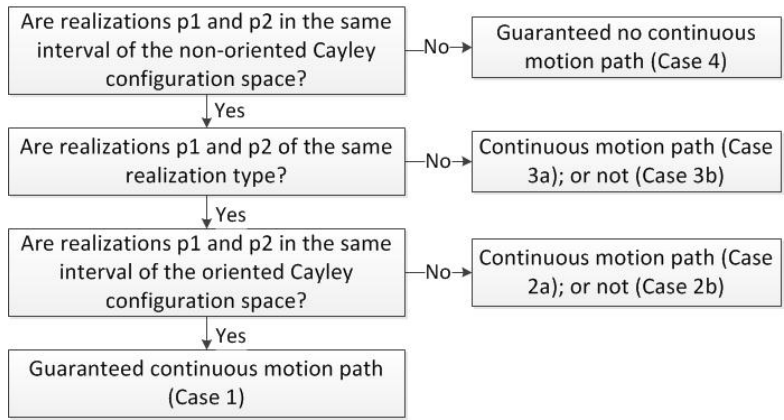}
\end{center}
\caption{Complete case analysis of continuous motion paths between two realizations $p_1$ and $p_2$. 
}
\label{fig:diagram}
\end{figure*}

In \cite{Sitharam2011a}, 
algorithms are given for obtaining an (oriented) Cayley configuration space 
for generic 1-dof tree-decomposable linkages. 
Here a \emph{generic linkage} means that no bar length is zero, all bars have distinct lengths and 
at most one pair of adjacent bars can be collinear in any realization. 
A realization of a generic linkage automatically satisfies the usual notion of genericity in the rigidity literature.
The paper \cite{Sitharam2011a} also shows that for generic 1-dof tree-decomposable linkages with low Cayley complexity, 
the number of continuous motion paths between two realizations is \emph{at most two}, 
and can be directly obtained from the oriented Cayley configuration spaces with complexity linear
in a natural, discrete measure of the length of the path.

The paper \cite{Sitharam2011b} gives the following theorem for recognizing low Cayley complexity linkages:
\begin{theorem}[Four-cycle Theorem]\label{the:four-cycle}
 A 1-dof tree-decomposable linkage has low Cayley complexity
if and only if every construction step of the underlying graph 
is based an adjacent pair of clusters, 
from a \emph{four-cycle} of clusters in the previously constructed graph. 
This yields an algorithm to recognize 1-dof tree-decomposable linkages with low Cayley complexity,
with time complexity quadratic in the number of construction steps of the underlying graph.
\end{theorem}


While judiciously chosen Cayley parameters  
shed light on many aspects of 1-dof tree-decomposable linkages,  
one persistent problem has been that 
a non-oriented Cayley interval, being a union of multiple oriented Cayley intervals, 
could correspond to multiple connected components of the realization space. 
On the other hand, although an oriented Cayley interval corresponds to a unique connected component, 
the mapping is not bijective, since that same connected component could contain more than one oriented interval \citep{sitharam2014beast}.
Figure \ref{fig:diagram} summarizes the different cases when 
determining existence of a continuous motion path between two realizations. 
There are two cases (2 and 3) where there may or may not exist a continuous motion path (a and b).
Previous software  
mentioned in Section \ref{sec:soa} 
generate continuous motion within a specified Cayley interval, 
or multiple segments of continuous motion, 
each corresponding to a different oriented Cayley interval with the same realization type. 
Thus they cannot consistently distinguish Case 2a from Case 2b, 
or Case 3a from Case 3b. 

The papers \cite{sitharam2014beast,Sitharam2011a} give a bijective representation to represent the realization space, 
by characterizing a \emph{canonical} method of picking a minimal set of non-edges, called a \emph{complete Cayley vector}, such that adding those non-edges as bars results in global rigidity.
The distance vectors on these non-edges (called \emph{complete Cayley distance vectors}) for each realization of the realization space form a \emph{canonical Cayley curve}, which bijectively represents the realization space.
In addition, the \emph{Cayley distance} between two Cartesian realizations is defined to be the Euclidean distance between their complete Cayley distance vectors,
and the \emph{Cayley distance} between two connected components is defined to be the minimum Cayley distance taken over all pair of realizations, one from each component. 



In this paper, 
we present a new software \texttt{CayMos} implementing these existing theoretical results. 
In Section \ref{sec:contributions}, 
we list the contributions of this paper in the form of \texttt{CayMos} functionalities. 
We briefly describe the algorithms implemented in \texttt{CayMos} in Section \ref{sec:algorithms}, 
and give the corresponding pseudocode in Section \ref{sec:pseudocode}. 
In Section \ref{sec:classes} we describe the software architecture.

\section{Contributions: CayMos functionalities}
\label{sec:contributions}

Based on the theoretical  contributions of \cite{Sitharam2011a,Sitharam2011b,sitharam2014beast}, 
we present a new software \texttt{CayMos} \citep{bib:caymos} (web-accessible at http://www.cise.ufl.edu/\~{}menghan/caymos/, source code and userguide available at http://code.google.com/p/caymos/), 
which implements efficient algorithmic solutions for the following: 

\noindent\textbf{(1)} Determining low Cayley complexity and generating Cayley configuration spaces for a given linkage,
implementing \cite{Sitharam2011a,Sitharam2011b}.

\noindent\textbf{(2)} Meaningfully visualizing the 
connected components of the realization space as projection of the canonical Cayley curves, 
implementing \cite{Sitharam2011a,sitharam2014beast}, addressing Issue (a) from Section \ref{sec:soa}.

\noindent\textbf{(3)} Generating the connected components of the realization space, 
and finding a continuous motion path between two given realizations, 
implementing \cite{Sitharam2011a,sitharam2014beast}, addressing Issue (b) from Section \ref{sec:soa}.

\noindent\textbf{(4)} Finding the realizations representing the shortest Cayley distance between two different connected components of the realization space, 
implementing \cite{sitharam2014beast}, addressing Issue (c) from Section \ref{sec:soa}.

In the following we will briefly introduce the functionalities provided by \texttt{CayMos}.

\smallskip

\noindent \textbf{Note}: \texttt{CayMos} screen-shots and movies have been used in the papers \cite{sitharam2014beast,Sitharam2011a} to illustrate definitions and theoretical results. 
However this manuscript provides the first description of \texttt{CayMos} functionalities and architecture.


\subsection{Determining low Cayley complexity and generating Cayley configuration spaces}\label{subsec:ccs}
\begin{figure*}[hbtp]
\begin{center}
\includegraphics[width=1.1\linewidth]{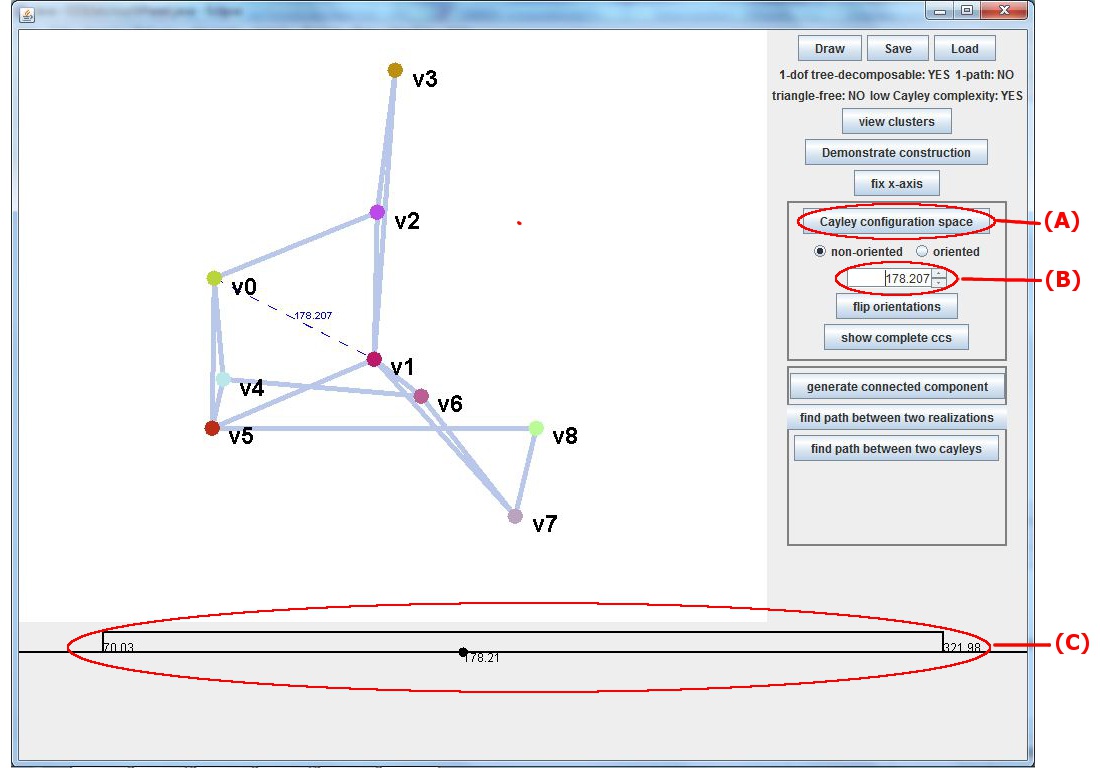}
\end{center}
\caption{Generating the non-oriented Cayley configuration space of the Limacon linkage using \texttt{CayMos}. 
(A) Button for generating the Cayley configuration space. 
(B) Spinner for specifying the length of the base non-edge and showing the corresponding realizations. 
(C) Intervals of the Cayley configuration space. The black dot denotes the Cayley configuration for the current realization. }
\label{fig:ccs}
\end{figure*}
%
As  Contribution (1), 
the user can generate the Cayley configuration spaces for 1-dof tree-decomposable linkages with low Cayley complexity. 
The following functionalities are provided: 

\noindent(\romannumeral 1) Determining whether the given linkage is 1-dof tree-decomposable with low Cayley complexity. 

\noindent(\romannumeral 2) Generating both the oriented and non-oriented Cayley configuration space(s)
for a linkage with low Cayley complexity, . 
The user can change the length of the chosen base non-edge to see corresponding realizations, as well as specify the realization type.

See Figure \ref{fig:ccs}  for the user interface. 

\subsection{Visualizing the connected components and finding continuous motion paths of the realization space}\label{subsec:connected_component}
\begin{figure*}[hbtp] 
\begin{center}
\includegraphics[width=1.05\textwidth]{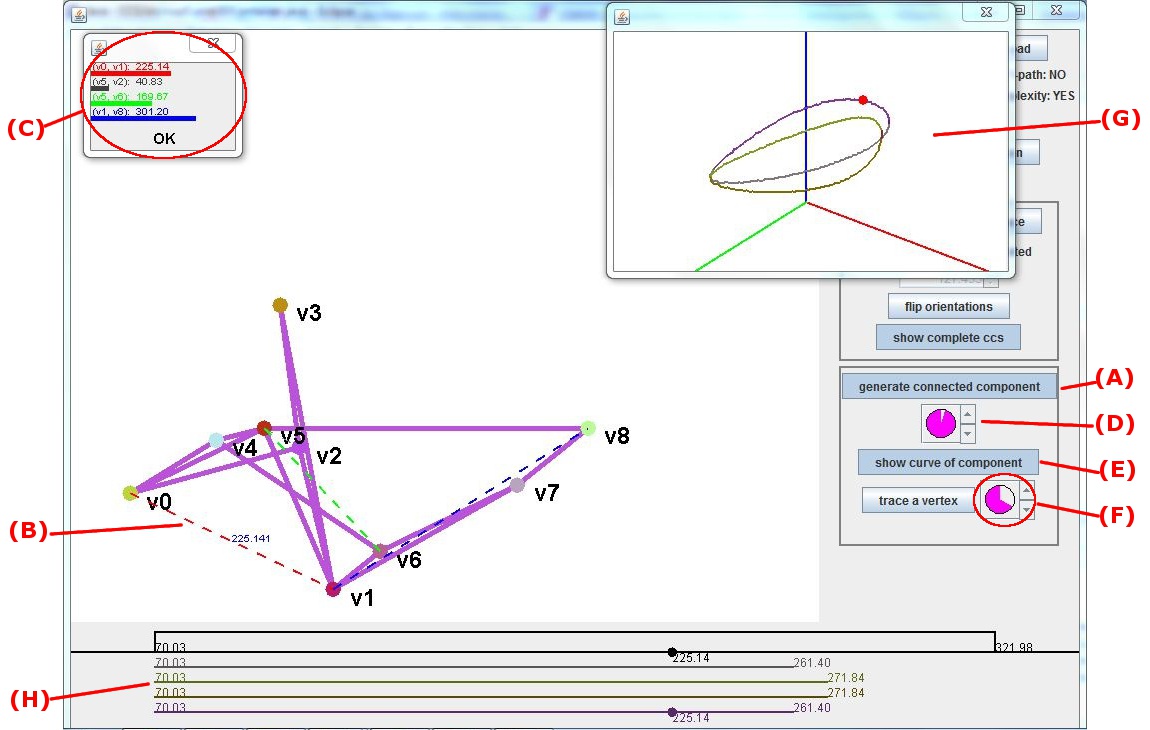}
\end{center}
\caption{Finding the connected components and showing corresponding canonical Cayley curves of 
the Limacon linkage using \texttt{CayMos}. 
(A) Button for generating the connected components. 
(B) The current realization, moving as the user traces the connected component. Dashed non-edges: non-edges in the complete Cayley vector. 
(C) Panel showing the complete Cayley distance vector for current realization. 
(D) Spinner for tracing the current connected component. 
(E) Button for showing the 3D projection of the corresponding canonical Cayley curve. 
(F) Spinner for navigating all connected components in the realization space. 
(G) The 3D projection of the canonical Cayley curve of the current component. 
The dot denotes the current realization and moves as the user traces the connected component. 
The curve is color-coded according to realization types. 
(H) The intervals of the oriented Cayley configuration spaces contained in the current connected component. }
\label{fig:components}
\end{figure*}
%

As Contribution (2) and (3),
the user can generate and visualize the connected components of the realization space,  
as well as find a continuous motion path between two realizations. 
The following functionalities are provided: 

\noindent(\romannumeral 1) Showing the non-edges in the complete Cayley vector of the linkage, 
as well as displaying the complete Cayley distance vector for the current realization. 
See Figure \ref{fig:components} (B) and (C). 


\noindent(\romannumeral 2) Generating all the connected components of the realization space,
as well as finding a continuous motion path (if one exists) between two realizations specified by the user.
See Figure \ref{fig:components} (A) and (F). 

\noindent(\romannumeral 3) Visualizing the connected component by showing the corresponding canonical Cayley curve, projected on three non-edges picked by the user from the complete Cayley vector. 
See Figure \ref{fig:components} (E), (G) and (D). 


\noindent(\romannumeral 4) Showing the curves traced out by vertices of the linkage in continuous motion. 

\subsection{Cayley distance between connected components}\label{subsec:distance}

As Contribution  (4), 
when the user tries to find a continuous motion path between two realizations in different connected components, 
\texttt{CayMos} will
find the two nearest realizations of these two components.

\section{\texttt{CayMos} software architecture and pseudocode}
\label{sec:architecture}

\subsection{Algorithms implemented in \texttt{CayMos}}
\label{sec:algorithms}

Several algorithms from our previous paper \cite{Sitharam2011a,Sitharam2011b,sitharam2014beast} are implemented in \texttt{CayMos}. 
Here we give a brief description of these algorithms.

\textbf{Four-cycle algorithm}: to determine whether a given 1-dof tree-decomposable linkage has low Cayley complexity, 
we implement the \emph{Four-cycle algorithm} \citep{Sitharam2011b},
which follows the construction of the linkage, and tests whether each construction step 
satisfies the condition given by the Four-cycle Theorem (see  Section \ref{sec:previous_cayley}).
For each of the $O(|V|)$ construction steps, we need to check $O(|V|)$ candidate base pairs of clusters, so
the overall time complexity is $O(|V|^2)$, 

\textbf{ELR algorithm}: 
to find the Cayley configuration space of a 1-dof tree-decomposable linkage with low Cayley complexity, 
we implement the \emph{ELR (extreme linkage realization) algorithm} from \cite{Sitharam2011a}.
The algorithm works by realizing all the \emph{extreme linkages} consistent with each realization type, 
and finding the intersection of all the candidate intervals of the base non-edge. 
As the problem of computing the Cayley configuration space is NP-hard \citep{Sitharam2011a},
the overall time complexity is exponential in $|V|$. 
The paper \cite{Sitharam2011a} also contains another algorithm called \emph{QIM (quadrilateral interval mapping)}, which also computes the Cayley configuration space, 
but only applies to a special subclass of linkages called \emph{1-path} 1-dof tree-decomposable linkages. 
The implementation of QIM algorithm is not included in \texttt{CayMos}.

\textbf{Continuous motion algorithm}: 
to find continuous motion paths and connected component(s) of the realization space, we implement the \emph{Continuous motion  algorithm} \citep{Sitharam2011a,sitharam2014beast}.
The algorithm works by traversing the intervals in the oriented Cayley configuration space via the common interval endpoints. 
As mentioned in Section \ref{sec:previous_cayley},
the time complexity is linear in the number of oriented Cayley configuration space endpoints along the continuous motion path or connected component.

\textbf{Closest pair algorithm}: 
to find the pair of ``closest'' realizations between two different connected components, 
we implement the \emph{Closest pair algorithm} from \cite{sitharam2014beast},
which samples both components and returns the pair of realizations with the smallest Cayley distance, 
which is computed using the complete Cayley distance vector (see Section \ref{sec:previous_cayley}).

\subsection{Major classes and architecture of \texttt{CayMos}}
\label{sec:classes}

Overall the backend of \texttt{CayMos} consists of two parts, with the following major classes. 

\subsubsection{1-dof tree-decomposable linkages and Cayley configuration spaces}

\begin{enumerate}

\item The \textsf{TDLinkage} class: represents a 1-dof tree-decomposable linkage. 

 Major Attributes:

\noindent --  \textsf{graph}: the underlying graph of the linkage. 

\noindent --  \textsf{barLengths}: the length of the bars of the linkage. 

\noindent --  \textsf{baseNonedge}: current base non-edge of the linkage.

\noindent --  \textsf{completeCayleyVector}: the complete Cayley vector for the current base non-edge.

\noindent --  \textsf{cayleyConfigurationSpace}: the Cayley configuration space on the current base non-edge. 

\smallskip
\noindent Major Methods:

\noindent-- \textsf{isLow()}: elaborated in Section \ref{sec:islow}.

\noindent-- \textsf{getCayleyConfigurationSpace()}: elaborated in Section \ref{sec:getcayleyconfigurationspace}.

\item The \textsf{Realization} class: represents a realization of a \textsf{TDLinkage} instance.

\noindent Major Attributes:

\noindent --  \textsf{tdLinkage}:
the corresponding 1-dof tree-decomposable linkage of the realization. 

\noindent --  \textsf{points}:
the 2D points in the realization for the vertices of the linkage. 

\smallskip

\noindent Major Methods:

\noindent --  \textsf{length(e:Edge)}: 
Returns the length of \textsf{e} in the realization, where \textsf{e} can be an edge or a non-edge.

\noindent --  \textsf{getCompleteCayleyDistanceVector}(): 
Returns the complete Cayley distance vector of the realization.
Calls \textsf{length()} for each non-edge in the \textsf{completeCayleyVector} field of \textsf{tdLinkage}, with time complexity $O(|V|)$.

\noindent --  \textsf{cayleyDistance(that:Realization)}: 
Returns the Cayley distance \citep{sitharam2014beast} between this realization and \textsf{that} with time complexity $O(|V|)$.

\item The \textsf{CayleyConfigurationSpace} and \textsf{OrientedCayleyConfigurationSpace} classes: represent the Cayley configuration space of a \textsf{TDLinkage} instance. Each \textsf{CayleyConfigurationSpace} contains multiple \textsf{OrientedCayleyConfigurationSpace} instances. 

\item The \textsf{OrientedInterval} class: represents an interval in an oriented Cayley configuration space. 

\noindent Major Attributes:

\noindent -- \textsf{upper} and \textsf{lower}:
the upper and lower endpoints of the interval.

\noindent --  \textsf{realizationType}:
the realization type of the \textsf{OrientedCayleyConfigurationSpace} containing the interval.

\noindent --  \textsf{nextIntervalUpper} and \textsf{nextIntervalLower}:
the two \textsf{OrientedInterval} instances sharing a common endpoint with this interval. 
They are immediately reachable from this interval in a continuous motion.

\end{enumerate}

\subsubsection{Continuous motion generation and representation}

\begin{enumerate}

\item The \textsf{ContinuousMotion} class: represents a continuous motion path between two \textsf{Realization} instances, or a connected component of a \textsf{TDLinkage} instance.

\noindent  Major Attributes:

\noindent -- \textsf{tdLinkage}: 
the corresponding 1-dof tree-decomposable linkage of the continuous motion. 

\noindent --  \textsf{orientedIntervals}: 
the list of \textsf{OrientedInterval} instances encountered along the continuous motion.

\smallskip

\noindent Major methods:

\noindent -- \textsf{findComponent()}, \textsf{findPath()} and \textsf{findAllComponents()}: elaborated in Section~\ref{sec:findpath}.

\noindent -- \textsf{findNearestRealizations()}: elaborated in Section~\ref{sec:findnearest}.

\noindent -- \textsf{getRealizations()}: 
Returns a list of realization samples along the continuous motion by sampling the \textsf{orientedIntervals} field. 
Uses a sampler object so that different ways of sampling can be chosen at runtime. 

\noindent --  \textsf{get3DCurve(f1,f2,f3)}: 
Returns a \textsf{Curve3D} instance representing the 3D projection of the continuous motion's corresponding Cayley curve
on \textsf{f1}, \textsf{f2} and \textsf{f3}, three non-edges from the complete Cayley vector of tdLinkage. 
The time complexity is linear in the size of the list returned by \textsf{getRealizations()}.

\item The \textsf{Curve3D} class: supports visualization of a  \textsf{ContinuousMotion} instance as a Cayley curve projected in 3D.

\end{enumerate}

\begin{figure*}[hbtp]
\begin{center}
\includegraphics[width=1.1\linewidth]{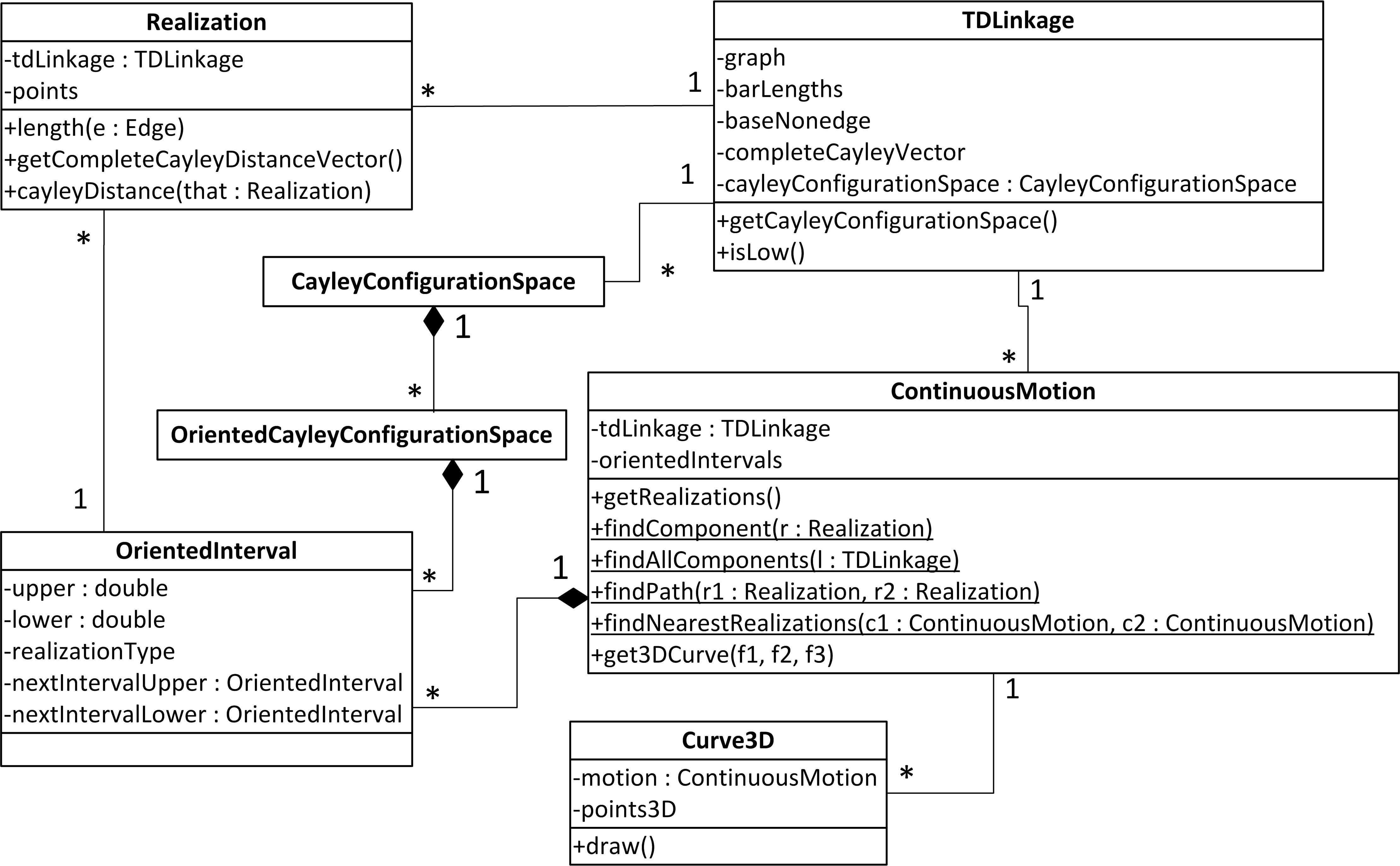}
\end{center}
\caption{UML diagram for major classes. }
\label{fig:uml}
\end{figure*}

Figure \ref{fig:uml} shows the relationships between the above classes.

\subsection{Major methods implementing algorithms in Section~\ref{sec:algorithms}}
\label{sec:pseudocode}

\subsubsection{ Method \textsf{isLow()}} 
\label{sec:islow}


The medthod implements the Four-cycle algorithm \citep{Sitharam2011b} sketched in Section \ref{sec:algorithms} to determine low Cayley complexity.
It also implements  Theorem 3 from \cite{sitharam2014beast} to find the complete Cayley vector
and store it in  the  \textsf{completeCayleyVector} field. 
Since both algorithms follow the construction of the linkage, we combine the implementation into one method. 
The  time complexity is $O(|V|^2)$. 

\smallskip
\noindent\textbf{Note:} in the current version of the software, the complete Cayley vector is implemented 
as in \cite{sitharam2014beast}, which is not minimal. 
The \emph{minimal complete Cayley vector} introduced in \cite{Sitharam2011a} will have dimension two for a large number of 1-dof tree-decomposable linkages. 

\smallskip

\lstset{language=Java} 

\lstset{
  language=Java,
  tabsize=2,
  basicstyle=\footnotesize\sffamily,
  breaklines=true
}

 Pseudocode for  \textsf{isLow()} :

\begin{lstlisting}[mathescape]
boolean isLow() {
	completeCayleyVector.add(baseNonedge);

	for (each construction step s) {
	// Construction Step s: adding two clusters s.c1 and s.c2, s.v $\triangleleft$ (s.v1, s.v2), s.v1 $\in$ s.c1, s.v2 $\in$ s.c2
		if (s is not directly based on the base non-edge) { 
			// find a valid pair of base clusters
			for (each Vertex w in the previously constructed graph sharing a cluster with both $s.v1$ and $s.v2$){
				Cluster c1 = the cluster containing s.v1 and w;
				Cluster c2 = the cluster containing s.v2 and w;
				if (validBasePairs.contains((c1,c2)){
					isLowStep = true;
					completeCayleyVector.add(nonedge (w, s.v));
					currentBasePairs.add((c1,c2));
				}
			}

			if (!isLowStep)  // does not have low Cayley complexity
				return false;

			for (Pair(c1,c2) in currentBasePairs){
				validBasePairs.add((s.c1, c1));
				validBasePairs.add((s.c2, c2));
			}
		}
		validBasePairs.add((s.c1,s.c2));
	}
	return true;
}
\end{lstlisting}

\subsubsection{ Method \textsf{getCayleyConfigurationSpace()}}
\label{sec:getcayleyconfigurationspace}
The method implements the ELR algorithm \citep{Sitharam2011a} sketched in Section \ref{sec:algorithms}
to generate the 
Cayley configuration space on the current base non-edge. 
As pointed out in \cite{Sitharam2011a}, computation of the Cayley configuration space is NP-hard, and this method can take time exponential in $|V|$.

\smallskip
 Pseudocode for  \textsf{getCayleyConfigurationSpace()} :

\begin{lstlisting}[mathescape]
CayleyConfigurationSpace getCayleyConfigurationSpace() {
	for (each construction step s) {
		for (each extreme linkage Realization r at step s) {
			distance = r.length(baseNonedge);
			for (each complete solution type t compatible with the partial solution type of r) 
				candidateEndpointLists[t].add(distance);
		}
	}
	
	for ( (SolutionType:t,List:l) : candidateEndpointLists) {
		sort(l);
		occs = new OrientedCaylyConfigurationSpace with SolutionType t;
		for (double cur : l) {
			double prev = the point before cur in l;
			double next = the point after cur in l;
			boolean P = realizable((prev + cur) / 2, t);
			boolean N = realizable((cur + next) / 2, t);
			if (!P && !N) {
				// cur is an isolated point
				occs.appendInterval(cur, cur); 
			} else if (P && !N) {
				// cur is end of interval
				occs.appendInterval(lastEndpoint, cur); 
				intervalStart = null;
			} else if (!P && N){
				// cur is start of interval
				intervalStart = cur; 
			} // else: cur is in middle of interval
		}
		this.cayleyConfigurationSpace.addOrientedCayleyConfigurationSpace(occs);
	}
	return this.cayleyConfigurationSpace;
}
\end{lstlisting}

\subsubsection{Methods \textsf{findPath(r1:Realiztion, r2:Realization)}, \textsf{findComponent(r:Realization)}, \textsf{findAllComponents(l:TDLinkage)} } \label{sec:findpath}
These methods implement the Continuous motion algorithm \citep{Sitharam2011a,sitharam2014beast} sketched in Section \ref{sec:algorithms}.
Method \textsf{findPath(r1,r2)} 
returns an instance of \textsf{ContinuousMotion}
representing the continuous motion path between realization \textsf{r1} and \textsf{r2}, 
which are realizations of the same linkage. 
Method \textsf{findComponent(r)} returns an instance of \textsf{ContinuousMotion}, 
which is the connected component in the realization space containing the realization \textsf{r}. 
Method \textsf{findAllComponents(l)}
returns a list of \textsf{ContinuousMotion} instances representing all connected components in the realization space of the linkage \textsf{l}. 

Both \textsf{findPath()} and \textsf{findComponent()}  
have time complexity linear in the number of oriented Cayley configuration space endpoints contained in the 
\textsf{ContinuousMotion} instance returned. 
Method \textsf{findAllComponents()} 
has time complexity linear in the total number of endpoints of all oriented Cayley configuration spaces \citep{sitharam2014beast}. 

\smallskip
 Pseudocode for \textsf{findComponent()} (\textsf{findPath()} is implemented similarly):
\begin{lstlisting}
static ContinuousMotion findComponent(Realization r) {
	startInt = the OrientedInterval containing the Cayley configuration of r;
	component.orientedIntervals.add(startInt);

	OrientedInterval curInt = startInt; 
	double endlf = curInt.lower;
	while (true) {
		curInt = (endlf == curInt.lower? curInt.nextIntervalLower:curInt.nextIntervalUpper)
		if (curInt == startInt) 
			return component;	
		endlf = (endlf == curInt.lower? curInt.upper:curInt.lower);
		component.orientedIntervals.add(curInt);
	}
}
\end{lstlisting}

The method  \textsf{findAllComponents()} is implemented by iterating over all intervals 
in every \textsf{OrientedCayeyConfigurationSpace} and 
calling \textsf{findComponent}.

%
%
%

\subsubsection{Method \textsf{findNearestRealizations(c1: ContinuousMotion, 
 c2: ContinuousMotion)}}\label{sec:findnearest}
 The method implements the Closest pair algorithm \citep{sitharam2014beast}
 sketched in Section \ref{sec:algorithms}.
The method returns the two nearest realizations between the two connected components 
\textsf{c1} and \textsf{c2} of the realization space of a linkage, 
using the Cayley distance measure. 
%
The time complexity $O(k^2|V|)$, 
where $k$ is the size of the list returned by \textsf{getRealizations()},
which is a  list of sample realizations in the connected component.

\medskip
\noindent \textbf{Note}: the sourcecode of \texttt{CayMos} is available at http://code.google.com/p/caymos/, 
and the software is web-accessible at http://www.cise.ufl.edu/\~{}menghan/caymos/.

\bibliographystyle{ACM-Reference-Format-Journals}
\bibliography{readings}

\end{document}